\documentclass[5p,times]{elsarticle}
\usepackage[utf8]{inputenc}
\usepackage[T1]{fontenc}
\usepackage{microtype}
\usepackage{booktabs}
\usepackage[english]{babel}
\usepackage{hyperref}
\usepackage{xr-hyper}
\usepackage{titlesec}

\usepackage{graphicx}

\usepackage{wrapfig}

\usepackage{amssymb}

\usepackage{tabularx}

\usepackage{array}

\usepackage{float}

\usepackage{dblfloatfix}

\usepackage{dirtytalk}

\usepackage{siunitx}

\usepackage{xcolor}

\usepackage{subcaption}

\usepackage{amsmath}

\usepackage{miller,soul}

\hypersetup{hidelinks}

\biboptions{sort&compress}

\journal{Materials Today Physics}

\begin{document}

\begin{frontmatter}

\title{A cylindrical sintering method for more realistic grain boundaries in nanocrystalline thin films}

%
\author[1,2]{Ankit Yadav\corref{cor1}}
\ead{ankityadav0811@gmail.com}
\cortext[cor1]{Corresponding author:}
\author[3]{Lucia Bajto\v sov\'a}
\author[3]{Miroslav Cieslar}
\author[1]{Jan Fikar}

\affiliation[1]{organization={Institute of Physics of Materials, The Czech Academy of Sciences}, addressline={Žižkova 22}, postcode={616 00}, city={Brno}, country={Czech Republic}}
\affiliation[2]{organization={Central European Institute of Technology, Brno University of Technology}, addressline={Purkyňova 123}, postcode={612 00},city={Brno}, country={Czech Republic}}
\affiliation[3]{organization={Charles University, Faculty of Mathematics and Physics}, addressline={Ke Karlovu 3},postcode={121 16}, city= {Prague}, country={Czech Republic}}

\begin{abstract}
Discrepancies between simulated and experimental mechanical properties
in molecular dynamics simulations of nanocrystalline metals typically
arise from the sample-construction method and the interatomic potential
choice. We introduce a cylindrical sintering method to generate
nanocrystalline aluminum thin-film samples with wider, more disordered
grain boundaries than the usual Voronoi tessellation method, while
maintaining deterministic control over grain size, shape, and
orientation. Cylindrical sintered samples are benchmarked against
hexagonal Voronoi references under identical conditions using both the
classical \textit{Pascuet15} MEAM and \textit{tabGAP} machine-learning
potentials. Cylindrical sintered samples consistently show lower
mechanical properties than hexagonal Voronoi samples due to their wider,
more disordered grain boundaries — an effect independent of the choice
of potential. Notably, changing the sample geometry and changing the
interatomic potential produce comparable, additive, and independent
shifts in predicted properties, highlighting that future molecular
dynamics studies must hold both variables fixed for meaningful
comparisons. Common neighbor and dislocation extraction analyses confirm
that deformation is dominated by grain-boundary-mediated plasticity.
Uniaxial tensile tests reveal an inverse Hall--Petch relationship for
both sample types and both potentials, with mechanical properties
decreasing monotonically as grain size reduces from 40.34 to 4.84\,nm.
The cylindrical sintering method offers a physically realistic,
geometrically controlled alternative that bridges idealized Voronoi
models and disordered experimental grain-boundary structures.

\end{abstract}


\begin{keyword}
	nanocrystalline aluminum, molecular dynamics,
	grain-boundary, inverse Hall--Petch
\end{keyword}

\end{frontmatter}

\section{Introduction}

Nanocrystalline (NC) metals, with mean grain sizes below approximately 100\,nm, exhibit a combination of high strength-to-weight ratio, enhanced hardness, and unusual deformation behavior that has motivated three decades of fundamental and applied research~\cite{GLEITER1989223,MEYERS2006427,KUMAR20035743}. In conventional polycrystalline metals, strength increases with decreasing grain size according to the Hall--Petch relation, $\sigma_{\mathrm{yield}} \propto d^{-1/2}$, as 
grain boundaries act as barriers to dislocation motion~\cite{Hall1951,
Petch1953} As the grain size is reduced into the deep-NC regime ($d \lesssim 20$\,nm), conventional dislocation-mediated plasticity is progressively replaced by grain-boundary-mediated processes such as boundary sliding, grain rotation, and the nucleation and absorption of partial dislocations at the boundaries themselves~\cite{Schiotz1998,VanSwygenhoven199811246,yamakov2004deformation}. The macroscopic manifestation of this crossover is the inverse Hall--Petch effect, in which the flow stress decreases rather than increases with decreasing grain size below a characteristic length scale of order 10\,nm. This crossover has been reproduced in MD simulations of nanocrystalline metals under ultra-high strain rates~\cite{Liu2025FeAl}, and a complementary two-phase description has been proposed for nanocrystalline Al thin films in which the grain-boundary region and grain interior are treated as mechanically distinct phases whose relative volume fractions set the overall grain-size dependence of strength~\cite{YADAV2024120084}.

Classical molecular dynamics (MD) has been the workhorse method for atomistic studies of NC plasticity, yet the absolute mechanical properties predicted by MD systematically differ from those measured experimentally on NC thin films~\cite{HAQUE20033053,Rajagopalan2011,BAJTOSOVA2022114688}. Two related contributions to this gap have been identified. The first is that grain boundaries produced by the most widely used sample-creation method---Voronoi tessellation~\cite{HIREL2015212}---are always very dense and almost ideal: planar, atomically sharp, and almost defect-free, in contrast to the wider, more disordered, and free-volume-rich boundaries observed by transmission electron microscopy in experimental NC films~\cite{BAJTOSOVA2022114688,Tristan2018}. Several routes address this limitation: introducing controlled porosity at grain boundaries to reintroduce natural free volume~\cite{YADAV2024120084}, using the melt--cool method to generate curved and atomically realistic boundaries directly from solidification~\cite{Yadav2025meltcool,MAHATA2019176,HOU2015256,Shibuta2017,HOU2015177}. A complementary strategy, analogous to powder metallurgy sintering, is to build the polycrystal by sintering together individual crystalline particles — perfect single-crystal grains of a chosen shape and orientation, placed in contact and allowed to consolidate under 
controlled temperature and pressure, with grain boundaries forming dynamically as atoms diffuse across the particle interfaces rather than 
being imposed as a geometric cut. Early MD work along these lines sintered two or three spherical Al nanoparticles of 3.2--4.4\,nm diameter to study the basic mechanics of neck formation and grain 
rotation~\cite{RAUT1998837}. Later studies investigated pairs of spherical Al nanoparticles of 4--10\,nm diameter to examine densification kinetics and grain-boundary formation~\cite{JIANG202092}, 
and extended the approach to graphene-nanoplatelet-reinforced Al matrix composites using particles of 3.24--6.48\,nm diameter, with sintering followed directly by tensile loading to evaluate the resulting mechanical properties~\cite{Huiping2019}. In all these cases, sintering was used to study either the sintering process itself or specific composite systems, rather than as a general-purpose route to constructing well-defined polycrystalline samples for deformation testing. The cylindrical sintering (CS) method introduced here takes this concept in a different direction.

We introduce the cylindrical sintering (CS) method, in which cylindrical grains are cut from a hexagonal Voronoi structure and sintered together to produce nanocrystalline thin-film samples with grain boundaries that are wider and more disordered than those of conventional Voronoi tessellation, while retaining full deterministic control over grain size, shape, and orientation. The use of cylindrical 
rather than spherical grains is motivated by the thin-film geometry: cylindrical grains naturally reproduce the columnar microstructure of physical vapour-deposited thin films and allow free surfaces to be 
introduced along the film normal direction, which is not possible with compact spherical-particle assemblies. The method is demonstrated here 
for nanocrystalline aluminum across grain sizes from 4.84 to 40.34\,nm. To verify that the observed grain-boundary characteristics and mechanical properties are intrinsic to the CS method rather than an artifact of any particular interatomic description, all simulations are performed with two potentials of qualitatively different character: the classical 
\textit{Pascuet15} MEAM potential~\cite{PASCUET2015229}, chosen for 
its established accuracy for Al elastic and defect properties, and the \textit{tabGAP} machine-learning potential~\cite{Fellman2025}, chosen for its broad DFT training database covering bulk, defect, surface, 
liquid, and cascade configurations of Al.

\section{Methods}
\label{sec:methods}

\subsection{Potential validation: single-crystal elastic constants}

The independent elastic constants $C_{11}$, $C_{12}$, and $C_{44}$ of fcc Al were computed in LAMMPS by applying small homogeneous strains to a single-crystal cell at 0\,K and 300\,K, and the polycrystalline elastic modulus $E_{sc}$, directional Young's moduli, Poisson's ratio $\nu$, and Zener anisotropy index $A = 2C_{44}/(C_{11}-C_{12})$ were derived following Holec \textit{et al.}~\cite{Holec2012}. Results for the \textit{Pascuet15} and \textit{tabGAP} potentials are compared against the experimental values of Vallin \textit{et al.}~\cite{Vallin1964} in Table~\ref{elasticcon}. Both potentials correctly reproduce the decreasing trend of elastic constants with increasing temperature and yield polycrystalline moduli at 300\,K within 12--14\% of experiment. The Zener anisotropy index approaches the experimental value of $A \approx 1.22$ at 300\,K for both potentials, confirming a reasonable description of the elastic anisotropy at the relevant simulation temperature.

The pretrained universal \textit{CHGNet} v0.3.0 potential~\cite{Deng2023CHGNet} was also evaluated as a third candidate. It yields $C_{44} \approx 0$\,GPa at 0\,K, at the boundary of the Born stability criterion~\cite{Born1954}, together with dramatic underestimation of the directional Young's moduli and a Poisson's ratio approaching 0.5, consistent with recent benchmarks reporting poor elastic-property predictions from pretrained \textit{CHGNet}~\cite{GaoWang2025elastic}. \textit{CHGNet} was therefore excluded from the subsequent nanocrystalline simulations.

\begin{table*}[t]
	\centering
	\caption{Single-crystal elastic constants $C_{ij}$, polycrystalline elastic modulus $E_{sc}$ (self-consistent method), directional Young's moduli $E_{\langle100\rangle}$, $E_{\langle110\rangle}$, $E_{\langle111\rangle}$, Poisson's ratio $\nu$, and Zener anisotropy index $A$ of fcc Al computed with the \textit{Pascuet15}, \textit{tabGAP}, and pretrained universal \textit{CHGNet}~v0.3.0~\cite{Deng2023CHGNet} potentials, compared against experimental values at 4\,K and 300\,K~\cite{Vallin1964}. \textit{CHGNet} was evaluated only at 0\,K.}
	\label{elasticcon}
	\begin{tabular*}{\textwidth}{@{\extracolsep{\fill}} l *{7}{c}}
		\hline \hline
		Parameter & \multicolumn{2}{c}{\textit{Pascuet15}} & \multicolumn{2}{c}{\textit{tabGAP}} & \textit{CHGNet} & \multicolumn{2}{c}{Experimental} \\
		\cline{2-3} \cline{4-5} \cline{6-6} \cline{7-8}
		& 0\,K & 300\,K & 0\,K & 300\,K & 0\,K & 4\,K & 300\,K \\
		\hline
		$C_{11}$ (GPa)                 & 112.64 & 74.46  & 100.79 & 84.62  & 60.64               & 116.3 & 107.3 \\
		$C_{12}$ (GPa)                 & 61.18  & 37.54  & 69.24  & 48.75  & 47.39               & 64.8  & 60.8  \\
		$C_{44}$ (GPa)                 & 44.96  & 28.43  & 36.85  & 26.32  & $-1.3\times10^{-3}$ & 30.9  & 28.3  \\
		$E_{sc}$ (GPa)                 & 93.78  & 61.93  & 71.54  & 60.34  & $2.2\times10^{-2}$  & 77.19 & 70.46 \\
		$E_{\langle100\rangle}$ (GPa)  & 69.57  & 49.30  & 44.40  & 48.98  & 19.06               & 69.93 & 63.32 \\
		$E_{\langle110\rangle}$ (GPa)  & 97.87  & 64.36  & 74.30  & 62.60  & $-5.2\times10^{-3}$ & 78.85 & 72.08 \\
		$E_{\langle111\rangle}$ (GPa)  & 113.22 & 71.67  & 95.80  & 68.99  & $-3.9\times10^{-3}$ & 82.35 & 75.56 \\
		$\nu$                          & 0.300  & 0.293  & 0.350  & 0.334  & 0.500               & 0.343 & 0.346 \\
		$A$                            & 1.747  & 1.54   & 2.336  & 1.467  & $-2.0\times10^{-4}$ & 1.20  & 1.217 \\
		\hline \hline
	\end{tabular*}
\end{table*}

\subsection{Sintering of cylindrical grains}
\label{sinteredcylindermethod}
\begin{figure*}[t]
	\centering
	\includegraphics[height=8cm]{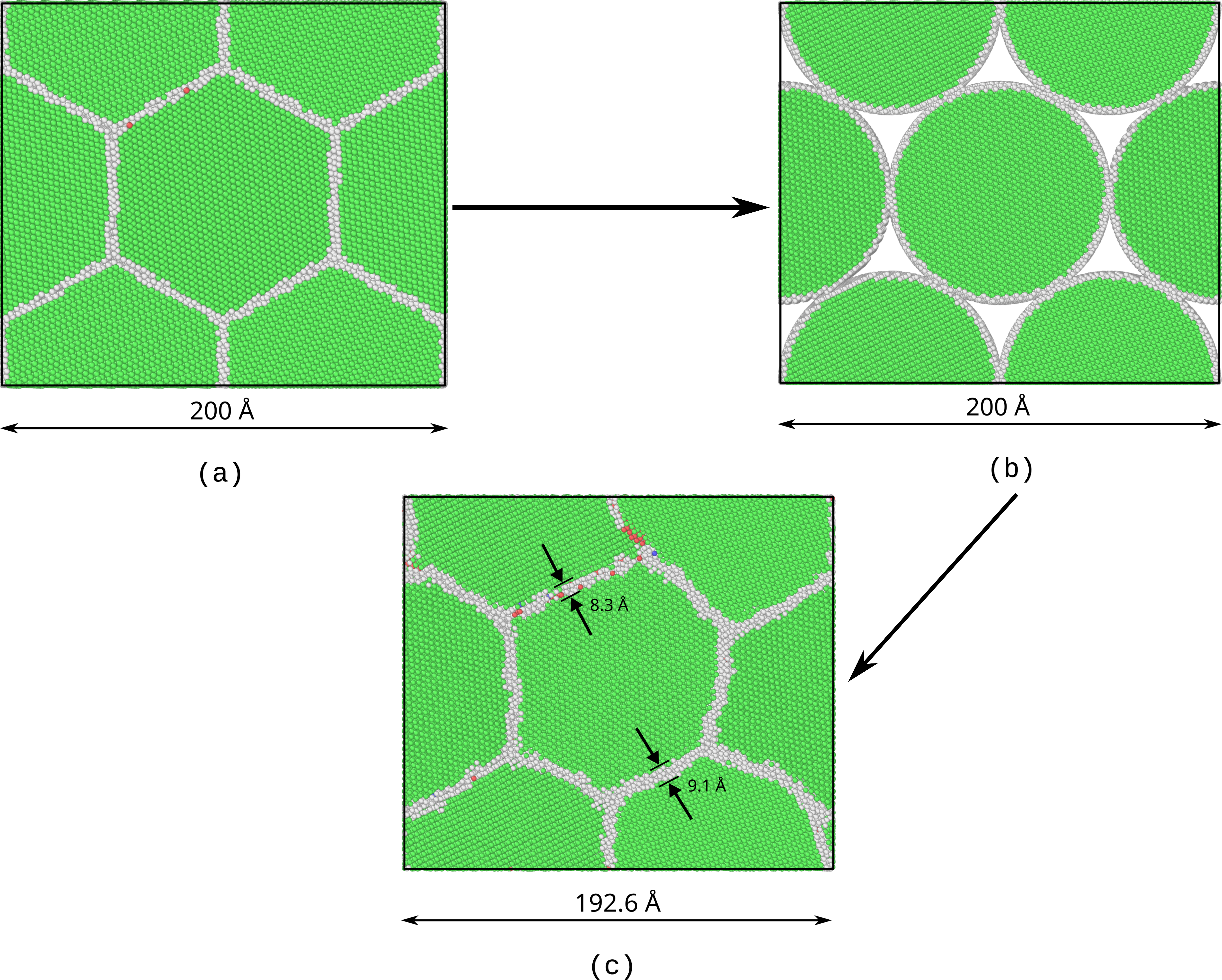}  
	\caption{Illustration of the cylindrical sintering construction procedure using the \textit{Pascuet15}~\cite{PASCUET2015229} MEAM potential. Top left: initial hexagonal Voronoi tessellation with a grain size of 10\,nm. Top right: cylindrical grains cut from the Voronoi structure, showing the inter-grain voids prior to sintering. Bottom: final sintered sample after densification and cooling, with a reduced grain size of 9.6\,nm and grain-boundary widths of 8.3\,\AA\ and 9.1\,\AA\ measured at two representative grain boundaries.}
	\label{fig:cylindersintered}
\end{figure*}

The cylindrical sintering (CS) method introduced here is inspired by Kadau \textit{et al.}~\cite{Kadau2004}, who sintered 32 spherical Al nanoparticles with three different diameters in a ratio of 2:3:4 and random orientations under 1--2\,GPa at 600\,K for 157.5\,ps, followed by relaxation at 300\,K. In the present work, cylindrical rather than spherical grains are used, motivated by the thin-film geometry: cylindrical grains naturally reproduce the columnar microstructure of physical vapour-deposited Al thin films and allow free surfaces to be 
introduced along the film normal direction. Grains were cut from hexagonal Voronoi structures with grain sizes ranging from 5 to 40\,nm, as illustrated in Fig.~\ref{fig:cylindersintered}, and populated with fcc Al crystals whose orientations were randomly assigned along all three crystallographic directions using Atomsk's \texttt{polycrystal} tool with the \texttt{random} keyword~\cite{HIREL2015212}. Ten independent 
configurations with different random orientations were generated per grain size, and the mechanical properties were subsequently averaged over these configurations.

The sintering step must strike a balance: sufficiently aggressive to close the inter-grain voids introduced by the cylinder-cutting procedure, but not so aggressive as to introduce bulk defects and dislocations that would compromise the subsequent deformation simulations. After systematic trials of pressure-controlled NPT and volume-controlled NVT approaches across a range of pressures, temperatures, and holding times, the optimal protocol was identified as NVT deformation to $0.93\,V_{\mathrm{eq}}$ at 600\,K for 200\,ps, which consistently produced void-free samples across all grain sizes from 10 to 40\,nm without significant bulk damage. The resulting 
internal pressures are 5.3, 5.1, 5.0, and 4.9\,GPa for 10, 20, 30, and 40\,nm, respectively. The complete protocol consisted of three stages: NPT equilibration at 600\,K for 10\,ps, deformation to 
$0.93\,V_{\mathrm{eq}}$ with NPT relaxation at 600\,K for 200\,ps, and cooling to 300\,K over 50\,ps. To satisfy periodic boundary conditions under fully random three-dimensional orientations, all samples were initially generated with a $z$-direction thickness of 20\,nm, from which a central 10\,nm slab was extracted after sintering 
for the tensile deformation simulations. The protocol was applied independently with both the \textit{Pascuet15}~\cite{PASCUET2015229} 
and \textit{tabGAP}~\cite{Fellman2025} potentials.


To isolate the effect of the cylindrical grain geometry on the sintered grain-boundary microstructure, the CS samples were compared against hexagonal Voronoi (HV) reference samples. The HV samples were generated from the same initial hexagonal Voronoi structure used to cut the cylinders, but without performing the cylinder-cutting step, and subjected to the identical three-stage sintering protocol in order to be directly comparable to the CS samples. The HV sample was constructed at the reduced box dimensions of the sintered CS sample, yielding a final grain size of 9.7\,nm, close to the 9.6\,nm of the CS sample. The CS sample exhibits rounded, curvilinear grain morphology reflecting the initial circular cross-sections, whereas the HV sample retains a more faceted, polygonal grain structure. Average grain-boundary thicknesses measured across both sample types and both potentials are summarized in Table~\ref{tab:gb_thickness_comparison}; the values are broadly comparable, suggesting that the initial grain geometry has a limited influence on the resulting grain-boundary thickness once sintering is complete.

\begin{table}[ht!]
	\centering
	\caption{Average grain-boundary thicknesses (\AA) measured on the same grain boundaries for the CS and HV samples, all constructed with fully random crystallographic orientations along all three axes, using the \textit{Pascuet15}~\cite{PASCUET2015229} and \textit{tabGAP}~\cite{Fellman2025} potentials.}
	\label{tab:gb_thickness_comparison}
	\small
	\setlength{\tabcolsep}{4pt} 
	\begin{tabular}{lccccc}
		\toprule
		\textbf{Sample} & \textbf{Grain} & \multicolumn{2}{c}{\textit{\textbf{Pascuet15}}} & \multicolumn{2}{c}{\textit{\textbf{tabGAP}}} \\
		& \textbf{size (nm)} & & & & \\
		\cmidrule(lr){3-4} \cmidrule(lr){5-6}
		& & \textbf{GB1 (\AA)} & \textbf{GB2 (\AA)} & \textbf{GB1 (\AA)} & \textbf{GB2 (\AA)} \\
		\midrule
		CS  & 9.6 & 8.3 & 9.1 & 7.3 & 9.8 \\
		HV & 9.7 & 8.4 & 7.9 & 7.0 & 9.7 \\
		\bottomrule
	\end{tabular}
\end{table}

\subsection{MD simulation setup and mechanical straining}

Molecular dynamics simulations were carried out using the LAMMPS software package~\cite{PLIMPTON19951} with a time step of 1\,fs. Prior to molecular dynamics, all polycrystalline samples were relaxed using the conjugate-gradient (CG) method with a relative energy tolerance of $10^{-1}$ to remove high atomic forces in the as-constructed configurations.

A preliminary strain-rate sensitivity study was performed on a sample containing four regular hexagonal grains with a grain size of 10\,nm ($200 \times 173.2 \times 100\,\text{\AA}^3$), constructed in the same way as the HV reference samples, using both the \textit{Pascuet15}~\cite{PASCUET2015229} and 
\textit{tabGAP}~\cite{Fellman2025} potentials at strain rates of 
$2\times10^{7}$, $2\times10^{8}$, $2\times10^{9}$, and 
$2\times10^{10}\,\mathrm{s}^{-1}$. A strain rate of 
$2\times10^{9}\,\mathrm{s}^{-1}$ was selected as the best compromise 
between computational cost and convergence toward the rate-independent 
regime and was used for all subsequent CS and HV simulations.

Prior to tensile loading, all samples underwent energy minimization with a relative tolerance of $10^{-8}$. Uniaxial tensile deformation was applied along the $x$-axis to 50\% strain at $2\times10^{9}\,\mathrm{s}^{-1}$, with zero stress maintained in the $y$ and $z$ directions, at 300\,K and zero pressure using a Nos\'e--Hoover thermostat and barostat~\cite{Nose10061984,Hoover1985}. The longitudinal and transverse engineering strains are defined as

\begin{equation}
    \label{eqex}
    \varepsilon_x = \frac{\Delta L_x}{L_{x0}}, \qquad
    \varepsilon_y = \frac{\Delta L_y}{L_{y0}}.
\end{equation}

The elastic modulus, ultimate tensile strength, and engineering yield stress were extracted from the stress--strain curves. Microstructural analysis was performed in OVITO~\cite{Stukowski2012} using Common Neighbor Analysis (CNA) and the Dislocation Extraction Algorithm (DXA). DXA signals within grain-boundary regions should not be interpreted as a measure of plastic activity, since the intrinsic disorder at grain boundaries produces apparent dislocation segments in the absence of plasticity-carrying dislocations; DXA is used here primarily to confirm that grain interiors remain free of dislocation segments throughout deformation.

\section{Results and discussion} 
\label{Resutlsanddiscussion}

\subsection{Strain-rate sensitivity}
\label{sec:strainrate}

The mechanical response of nanocrystalline metals is sensitive to the applied strain rate~\cite{Meyers_Chawla_2008}. In bulk metallic 
materials this dependence is often expressed through the power-law relation
\begin{equation}
\label{eq:powerlaw}
\sigma_{\mathrm{yield}} = K \dot{\varepsilon}^{\,n},
\end{equation}
where $n$ is the strain-rate sensitivity exponent, typically 0.02--0.2 for metals~\cite{KASSNER20001,Rosen1999,GAMBIRASIO2016231}. Several MD studies of fcc nanocrystalline metals have reported positive strain-rate sensitivity together with transitions between qualitatively distinct deformation regimes~\cite{Cabral2025,PastorAbia2011,
OUYANG2024107752}, motivating a systematic test across strain rates from $2\times10^{7}$ to $2\times10^{10}\,\mathrm{s}^{-1}$ on a sample 
with a mean grain size of 7.8\,nm, performed independently with both the \textit{Pascuet15} and \textit{tabGAP} potentials.

Fitting Eq.~\ref{eq:powerlaw} to the ultimate tensile stresses across all four strain rates gives $n = 0.08$ for both \textit{Pascuet15} and \textit{tabGAP}, consistent with the range reported for nanocrystalline Al in the literature~\cite{Cabral2025,PastorAbia2011}. It should be noted that $n$ is known to depend on grain size in nanocrystalline metals~\cite{Cabral2025,PastorAbia2011}, the value reported here is specific to the 10\,nm grain size used in this study.

\subsection{Mechanical behavior of CS and HV samples}
\label{sec:mechbehaviour}
The two-phase model developed in our previous work~\cite{YADAV2024120084} is applied here to fit the simulated mechanical-property data for the CS and HV samples across the full grain-size range investigated, from 5 to 40\,nm. The model is used here in parallel with two distinct potentials, the \textit{Pascuet15}~\cite{PASCUET2015229} MEAM potential and the \textit{tabGAP}~\cite{Fellman2025} machine-learning potential, in order to isolate any potential-induced offset in the extracted grain-interior and grain-boundary parameters. The resulting fit parameters $E_g$, $E_b$, $t$, $\sigma_g$, $\sigma_b$, $\sigma_{0.2_g}$, and $\sigma_{0.2_b}$ are summarized in Table~\ref{tab:twophase_sintered} for both potentials and both sample-construction methods.

\begin{table}[ht!]
	\caption{Fitted parameters of the two-phase model for the CS and HV samples, simulated using both the \textit{Pascuet15}~\cite{PASCUET2015229} MEAM potential and the \textit{tabGAP}~\cite{Fellman2025} machine-learning potential. $E_g$ and $E_b$ denote the elastic moduli of the grain interior and grain-boundary respectively; $\sigma_g$, $\sigma_b$, $\sigma_{0.2_g}$, and $\sigma_{0.2_b}$ represent the ultimate tensile strength and engineering yield stress contributions from the grain interior and grain-boundary respectively; $t$ is the grain-boundary thickness.}
	\centering
	\small
	\setlength{\tabcolsep}{2pt} 
	\begin{tabular*}{\columnwidth}{@{\extracolsep{\fill}} lcccc}
		\hline \hline
		Parameter & \multicolumn{2}{c}{\textit{\textbf{Pascuet15}}} & \multicolumn{2}{c}{\textit{\textbf{tabGAP}}} \\
		
		\cline{2-3} \cline{4-5}
		& CS & HV & CS & HV \\
		\hline
		$E_g$ (GPa)            & 57.9 & 63.9 & 52.2 & 60.3 \\
		$E_b$ (GPa)            & 29.6 & 34.1 & 17.6 & 32.2 \\
		$t$ (nm)               & 1.03  & 1.94  & 0.44  & 2.25 \\
		$\sigma_g$ (GPa)       & 2.46  & 2.77  & 1.90  & 2.42 \\
		$\sigma_b$ (GPa)       & 7.2$\times10^{-5}$ & 0.11 & 8.39$\times10^{-3}$ & 4.9$\times10^{-2}$ \\
		$\sigma_{0.2_g}$ (GPa) & 2.28 & 2.60 & 1.85 & 2.38 \\
		$\sigma_{0.2_b}$ (GPa) & 4.74$\times10^{-6}$ & 3.42$\times10^{-5}$ & 2.38$\times10^{-9}$ & 5.88$\times10^{-5}$ \\
		\hline \hline
	\end{tabular*}
	\label{tab:twophase_sintered}
\end{table}

\begin{figure*}[t]
	\centering
	\begin{subfigure}{0.5\textwidth}
		\centering
		\includegraphics[height=5.5cm]{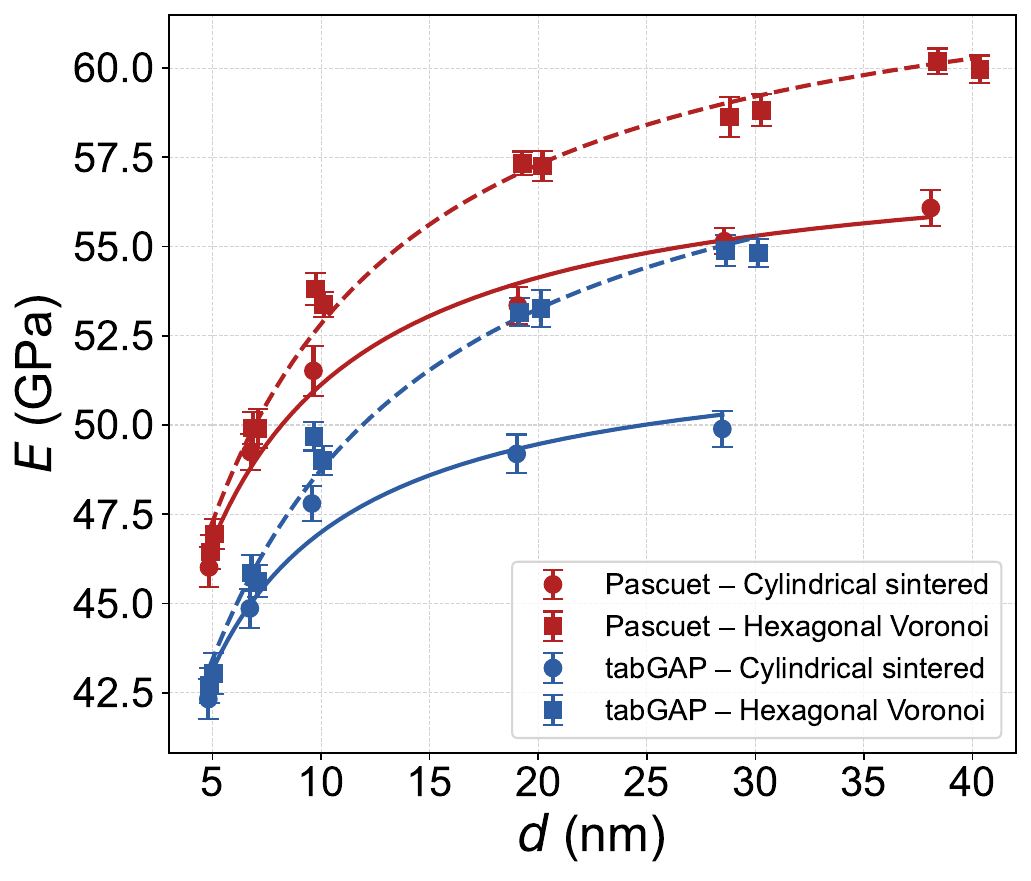}
		\caption{}
		\label{E_cylindrical}
	\end{subfigure}%
	\begin{subfigure}{0.5\textwidth}
		\centering
		\includegraphics[height=5.5cm]{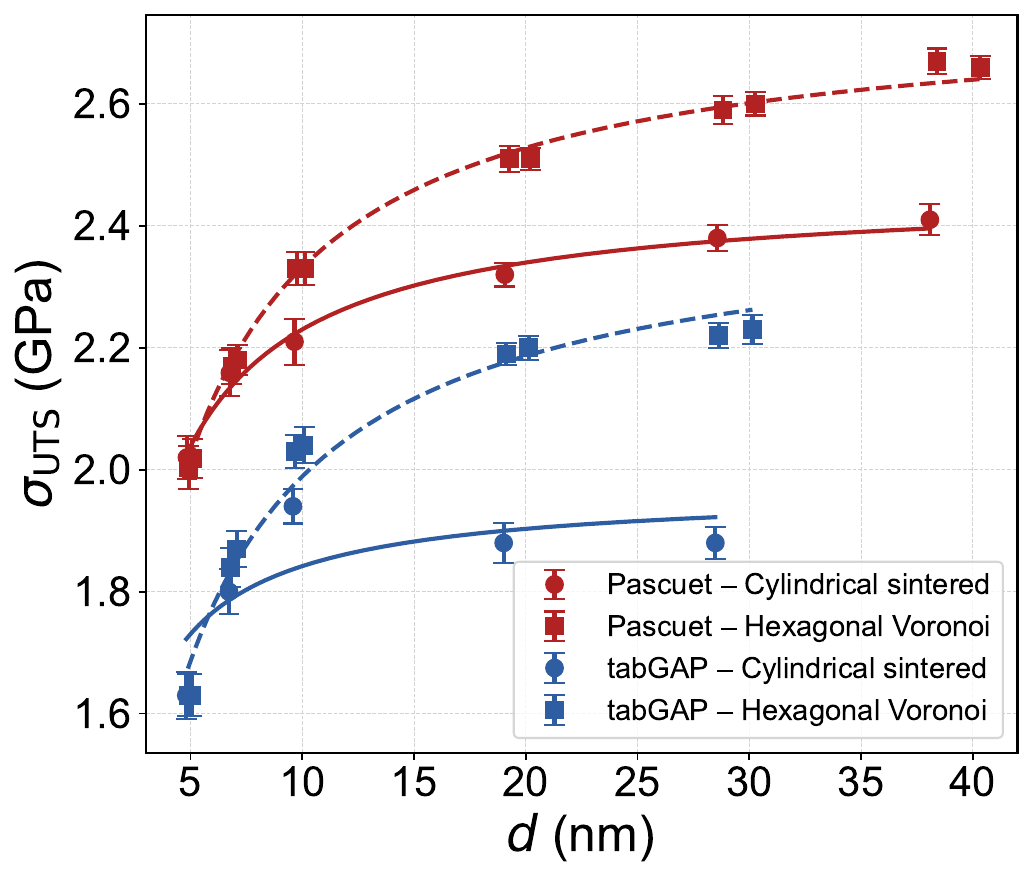}
		\caption{}
		\label{UTS_cylindrical}
	\end{subfigure}
	
	\vspace{0.3cm} 
	
	\begin{subfigure}{0.5\textwidth}
		\centering
		\includegraphics[height=5.5cm]{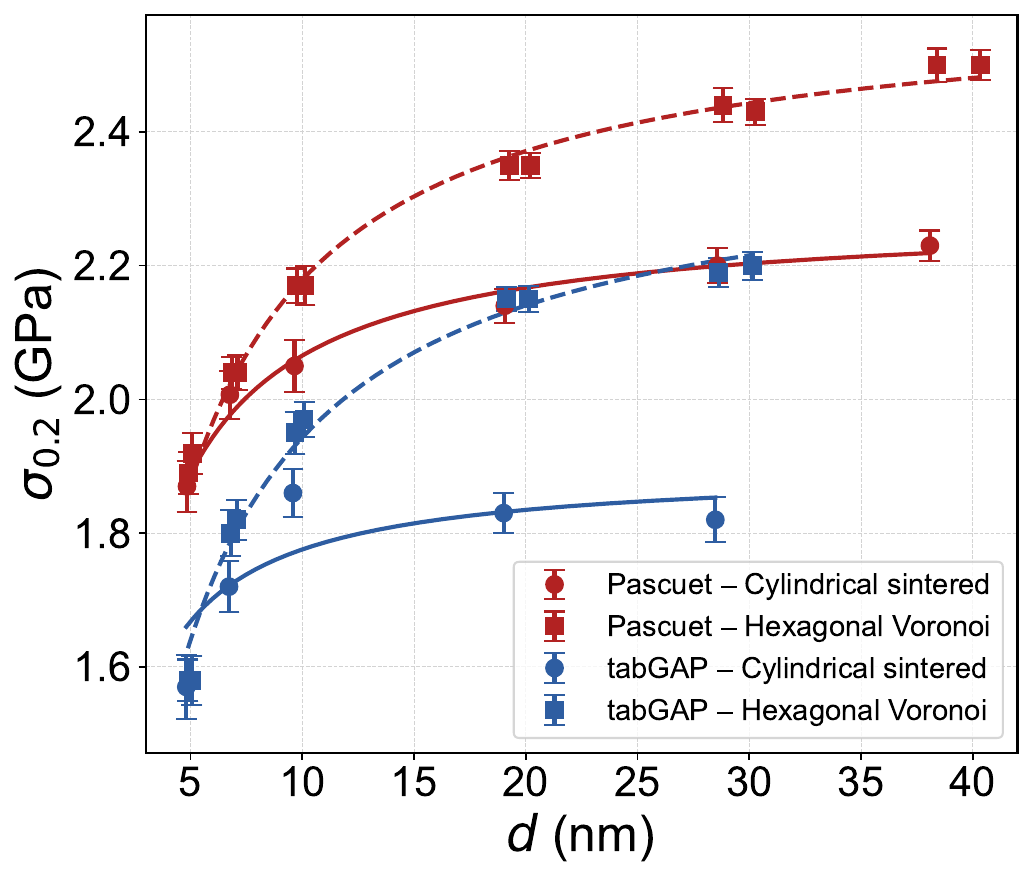}
		\caption{}
		\label{sigma0.2_cylindrical}
	\end{subfigure}
	
	\caption{Size dependence of mechanical properties of nanocrystalline Al samples constructed with fully random crystallographic orientations along all three axes. Sample geometries: cylindrical sintered (CS, squares) and hexagonal Voronoi (HV, circles). Interatomic potentials: \textit{Pascuet15}~\cite{PASCUET2015229} (red) and \textit{tabGAP}~\cite{Fellman2025} (blue): (a) elastic modulus $E$, (b) ultimate tensile strength $\sigma_{\mathrm{UTS}}$, (c) engineering yield stress $\sigma_{0.2}$. Solid lines indicate the two-phase model fit for CS samples and dashed lines for HV samples. The error bars are standard deviations over 10 differently oriented configurations.}
	\label{mechanical_cylindrical}
\end{figure*}

The mechanical properties of the CS and HV samples are presented as a function of grain size $d$ in Figs.~\ref{E_cylindrical},~\ref{UTS_cylindrical}, and~\ref{sigma0.2_cylindrical}. The dual-potential design allows two distinct comparisons to be made: the influence of the sample-construction method (CS vs HV) for each potential separately, and the influence of the interatomic potential (\textit{Pascuet15} vs \textit{tabGAP}) for each sample type separately. In all three figures, squares indicate CS samples and circles indicate HV samples (sample comparison by symbol shape), while red indicates the \textit{Pascuet15} potential and blue indicates the \textit{tabGAP} potential (potential comparison by color). The solid and dashed lines represent the corresponding two-phase model fits.

For both potentials, the CS samples consistently yield lower mechanical properties than the HV samples across the full grain-size range. For the \textit{Pascuet15} potential, the elastic modulus of the CS samples ranges from 46 to 56\,GPa, while the HV samples reach 47 to 60\,GPa across grain sizes from 5 to 40\,nm. The ultimate tensile strength and engineering yield stress show the same offset: the CS samples reach $\sigma_{\mathrm{UTS}}$ of 2.02--2.41\,GPa and $\sigma_{0.2}$ of 1.87--2.23\,GPa, compared to 2.00--2.67\,GPa and 1.89--2.50\,GPa, respectively, for the HV samples. A similar offset is observed for the \textit{tabGAP} potential, where the CS samples yield 42--51\,GPa compared to 43--56\,GPa for the HV samples in elastic modulus, 1.63--1.88\,GPa vs 1.63--2.22\,GPa in $\sigma_{\mathrm{UTS}}$, and 1.57--1.82\,GPa vs 1.58--2.19\,GPa in $\sigma_{0.2}$. The two-phase model parameters in Table~\ref{tab:twophase_sintered} confirm this trend: for the \textit{Pascuet15} potential, $E_g=57.9$\,GPa for CS compared to $E_g=63.9$\,GPa for HV, with the grain-boundary elastic modulus $E_b$ also lower at 29.6\,GPa for CS compared to 34.1\,GPa for HV. The \textit{tabGAP} potential shows the same pattern, with $E_g=52.2$\,GPa for CS versus $E_g=60.3$\,GPa for HV, and a markedly larger difference in the grain-boundary elastic modulus, $E_b=17.6$\,GPa for CS compared to $E_b=32.2$\,GPa for HV. The grain-boundary thickness $t$ is also smaller for CS than for HV with both potentials: for \textit{Pascuet15}, $t=1.03$\,nm for CS compared to $t=1.94$\,nm for HV, and for \textit{tabGAP}, $t=0.44$\,nm for CS compared to $t=2.25$\,nm for HV. This combination of a thinner but considerably softer grain boundary in the CS samples reflects the more disordered grain boundaries produced by the sintering of cylindrical grains, as discussed in Section~\ref{sinteredcylindermethod}.

A closer look at the two-phase model parameters in Table~\ref{tab:twophase_sintered} reveals that the lower mechanical properties of the CS samples cannot be explained by the grain-boundary thickness $t$ alone. For both potentials, the CS samples have a thinner grain boundary than the HV samples, yet they consistently exhibit lower mechanical properties. The explanation lies in the combined effect of the grain-boundary stiffness and the grain-interior modulus: the CS samples have both a softer grain boundary ($E_b$ lower than for HV) and a softer grain interior ($E_g$ lower than for HV), and these reductions are not compensated by the thinner grain-boundary layer. The grain-boundary strength contributions $\sigma_b$ and $\sigma_{0.2_b}$ are very small for all four cases ($\sigma_b \lesssim 10^{-1}\,\mathrm{GPa}$ and $\sigma_{0.2_b} \lesssim 10^{-4}\,\mathrm{GPa}$), indicating that the grain boundaries carry essentially no tensile load, so the macroscopic strength is governed by the grain interior while the macroscopic stiffness reflects the combined contributions of both phases.

Taken together, these observations demonstrate that the two-phase model parameters $E_g$, $E_b$, and $t$ together provide a more complete picture of the grain-boundary contribution to the mechanical properties than any single parameter alone. The interplay between grain-boundary thickness and stiffness gives the apparently unexpected result that thinner grain boundaries do not necessarily yield stiffer overall mechanical properties; rather, what matters is the combined effect of how thick the boundary is, how soft it is, and how stiff the surrounding grain interior remains.

Comparing the two interatomic potentials for each sample type, the \textit{Pascuet15} potential consistently predicts higher mechanical properties than the \textit{tabGAP} potential across all sample geometries and grain sizes. For the CS samples, the elastic modulus from \textit{Pascuet15} (46--56\,GPa) is systematically higher than that from \textit{tabGAP} (42--51\,GPa). A similar offset is observed for the HV samples, where \textit{Pascuet15} yields 47--60\,GPa compared to 43--56\,GPa for \textit{tabGAP}. This systematic offset is consistent with the higher elastic constants predicted by \textit{Pascuet15} at 300\,K (Table~\ref{elasticcon}). Since the same 
trend is reproduced independently for both CS and HV geometries, the potential-induced offset is unlikely to be an artifact of any particular 
sample-construction method; rather, it reflects intrinsic differences in how each potential describes the elastic response of the Al grain 
interior and grain boundary.

\subsection{Flow stress}
\label{sec:flowstress}
While the elastic modulus, ultimate tensile strength, and engineering yield stress are extracted from the well-defined elastic and yield regions of the stress--strain curve, the flow stress is of particular interest because some studies report a Hall--Petch transition in flow stress at grain sizes where other mechanical properties still follow inverse Hall--Petch behavior~\cite{Wang2025}. Extracting a reliable flow stress from the present simulations is, however, complicated by 
geometric instabilities that develop at larger grain sizes, as described below.

\begin{figure*}[t]
	\centering
	\begin{subfigure}{0.4\textwidth}
		\centering
		\includegraphics[height=5cm]{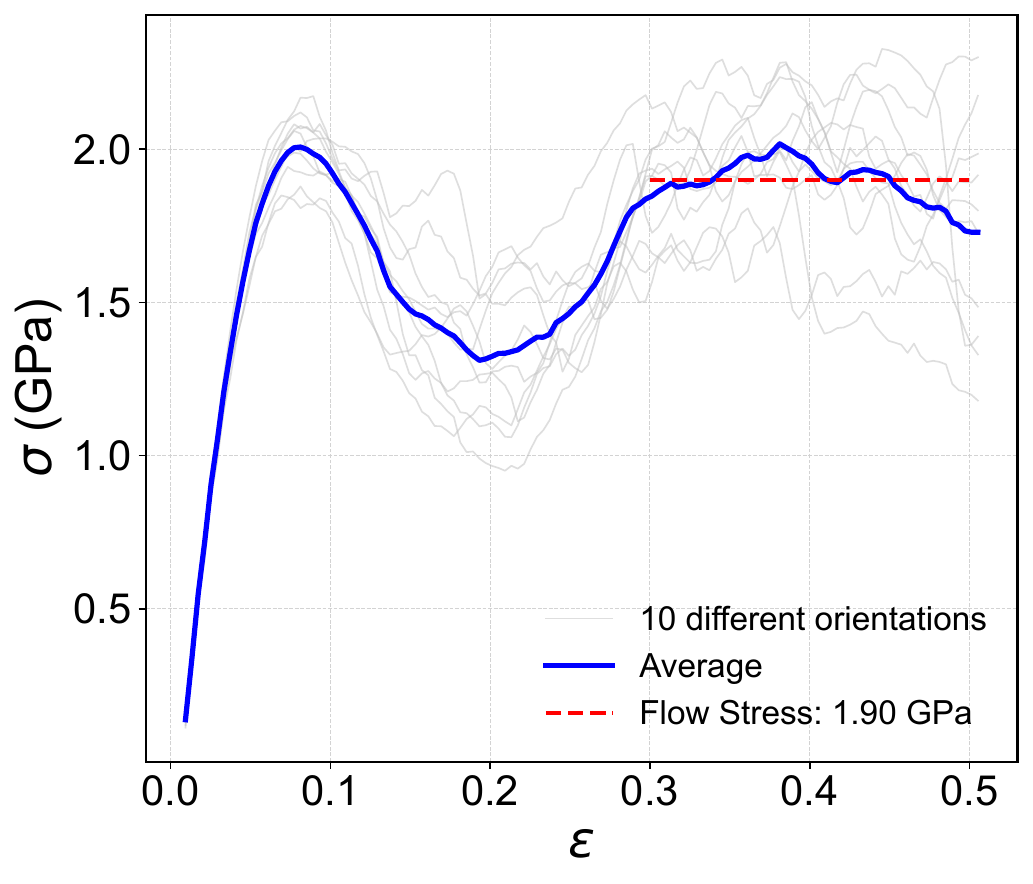}
		\caption{4.84\,nm}
		\label{fig:ss_0_5x}
	\end{subfigure}%
	\begin{subfigure}{0.4\textwidth}
		\centering
		\includegraphics[height=5cm]{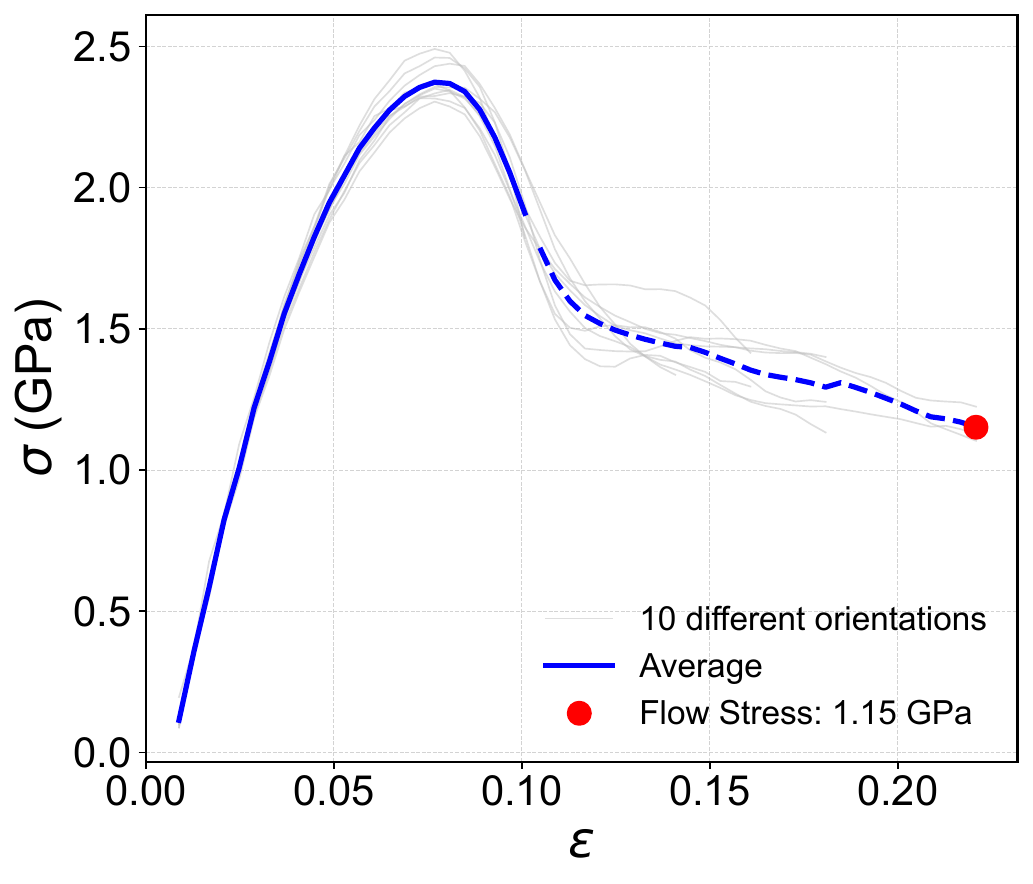}
		\caption{28.57\,nm}
		\label{fig:ss_3x}
	\end{subfigure}
	
	\caption{Stress--strain response of cylindrical sintered (CS) samples computed using the \textit{Pascuet15}~\cite{PASCUET2015229} MEAM potential, with fully random crystallographic orientations along all three axes. Simulations were performed for average grain sizes of 4.84, 6.76, 9.63, 19.06, and 28.57\,nm, each averaged over 10 random orientation configurations. Individual orientation-dependent curves are shown in gray, while the blue curve represents the mean stress--strain response. The 6.76 and 9.63\,nm cases closely resemble the 4.84\,nm response shown in panel (a) and are therefore not shown separately; likewise, the 19.06\,nm case closely resembles the 28.57\,nm response shown in panel (b) and is omitted for brevity.}
	\label{fig:stress_strain_all}
\end{figure*}

To account for the inherent variability introduced by crystallographic orientation, all stress--strain curves presented in this subsection were obtained by averaging over ten independent simulations performed with fully random grain orientations along all three axes for each grain size, for the cylindrical sintered (CS) samples simulated using the \textit{Pascuet15}~\cite{PASCUET2015229} MEAM potential. The individual orientation-dependent responses are shown as gray curves in Fig.~\ref{fig:stress_strain_all}, while the blue curve represents the mean stress--strain response. The plastic region of the individual orientation configurations is inherently noisy and exhibits significant stress fluctuations, making it difficult to extract a reliable flow stress from any single curve. Averaging over ten configurations smooths these fluctuations and produces a mean response with a sufficiently stable plastic plateau from which a representative flow stress can be meaningfully extracted.

For all grain sizes investigated, the stress--strain curves exhibit a well-defined yield peak. This is most likely a consequence of the 
absence of pre-existing free dislocations in the sintered samples: before plastic flow can begin, dislocations must first be nucleated 
from the grain boundaries under the applied stress, which requires a significant stress concentration and produces a pronounced load drop once the first dislocation sources become active.

In the smaller grain-size samples ($d \leq 9.63\,\mathrm{nm}$), the post-yield plastic region stabilizes into a relatively flat plateau across all ten orientation configurations. The flow stress for these samples was extracted as the statistical mean stress over the strain range $0.3 \leq \varepsilon \leq 0.5$. For the 4.84\,nm sample this is indicated in Fig.~\ref{fig:stress_strain_all}\subref{fig:ss_0_5x} as a horizontal dashed red line at $1.90\,\mathrm{GPa}$. The same procedure applied to the 6.76 and 9.63\,nm samples, not shown here as their stress--strain response closely resembles the 4.84\,nm case, yields flow stresses of $1.95\,\mathrm{GPa}$ and $1.87\,\mathrm{GPa}$, respectively.

For the larger grain sizes ($d = 19.06$ and $28.57\,\mathrm{nm}$), progressive necking prevents the formation of a stable plateau, and the flow stress was instead taken as the last point on the averaged stress--strain curve prior to complete failure. For the 28.57\,nm 
sample this is indicated in Fig.~\ref{fig:stress_strain_all}\subref{fig:ss_3x} as a red dot at $1.15\,\mathrm{GPa}$. The same procedure applied to the 19.06\,nm sample, not shown here as its response closely esembles the 28.57\,nm case, yields a flow stress of $1.19\,\mathrm{GPa}$. These values should 
be treated as lower-bound estimates of the true plastic resistance due to the geometric softening introduced by necking.

For these larger grain-size samples, a qualitatively distinct behavior emerges in the plastic regime, as shown in Fig.~\ref{fig:stress_strain_all}\subref{fig:ss_3x}. Following the yield peak, a subset of orientation configurations exhibits a sharp and 
progressive decline in stress, in some cases reaching values close to zero, rather than stabilizing at a finite flow stress. This does not reflect a change in the intrinsic plastic resistance of the material, but instead arises from geometric instability: strain localizes 
preferentially at grain boundaries, leading to pronounced necking, a severe reduction in the effective load-bearing cross-sectional area, and ultimately complete fracture in the most extreme cases.

\begin{figure*}[t]
	\centering
	\includegraphics[width=0.85\textwidth]{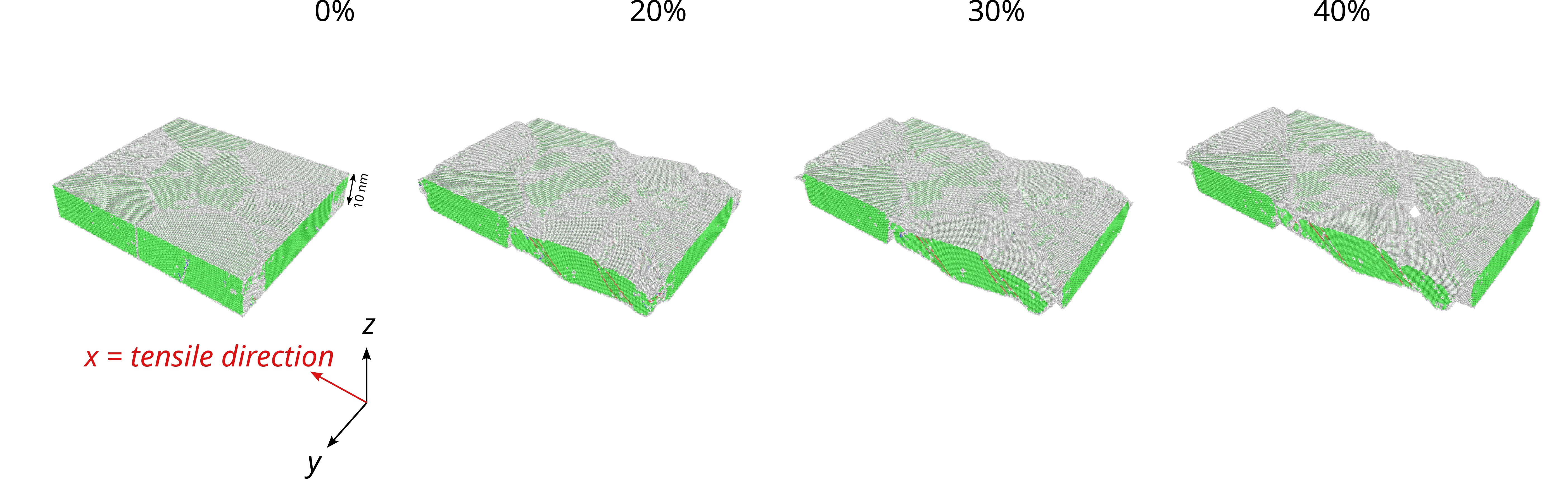}
	\caption{Microstructural evolution of the cylindrical sintered (CS) sample with an average grain size of 28.57\,nm at tensile strains of 0\%, 20\%, 30\%, and 40\%, simulated using the \textit{Pascuet15}~\cite{PASCUET2015229} MEAM potential and visualized via Common Neighbor Analysis (CNA) as implemented in  OVITO~\cite{Stukowski2012}. Green and white atoms represent crystalline and disordered grain-boundary atoms, respectively. The sample has free surfaces in the $z$-direction (film normal) with a slab thickness of 10\,nm; tensile loading is applied along the $x$-axis and the $y$-direction is unconstrained. Pronounced necking and strain localization are visible from 20\% strain onward, illustrating the geometric instability that complicates flow stress extraction at this grain size.}
	\label{fig:necking_cna}
\end{figure*}

This is clearly evidenced in Fig.~\ref{fig:necking_cna}, which presents 
CNA visualizations of the 28.57\,nm sample at engineering strains of 0\%, 20\%, 30\%, and 40\%. Pronounced necking is already apparent at 
20\% strain and intensifies progressively, with complete structural failure occurring by 40\% strain in several orientation configurations. 
The smaller grain sizes (9.63 and 19.06\,nm) show qualitatively similar but progressively less severe localization, with the 9.63\,nm sample deforming relatively homogeneously up to 50\% strain.

Including post-necking and post-fracture data in the statistical average would artificially deflate the mean flow stress, as the near-zero stress values recorded after fracture carry no physically meaningful information about the material's plastic resistance. To avoid this, stress--strain data recorded after the onset of severe necking or fracture were excluded from the averaging procedure for the affected orientation configurations. For these larger grain-size cases, the averaging window was restricted to the stable region immediately following the yield peak and prior to the onset of geometric failure, rather than the standard $0.3 \leq \varepsilon \leq 0.5$ range. Consequently, the flow stress values reported for $d = 19.06$ and $28.57\,\mathrm{nm}$ should be treated as lower-bound estimates of the true steady-state flow stress, or as indicators of the onset of plastic instability, rather than fully converged steady-state values.

This size-dependent transition from stable homogeneous plastic flow to localized necking and fracture is physically consistent with the grain-size dependence of deformation mechanisms in nanocrystalline metals. At smaller grain sizes, the high grain-boundary volume fraction promotes distributed GB-mediated plasticity, suppressing strain localization and stabilizing plastic flow. As grain size increases, intragranular dislocation activity becomes more prominent, deformation becomes less uniformly distributed, and the sample becomes increasingly susceptible to macroscopic strain localization and geometric instability under tensile loading.

To correct for the geometric change in cross-sectional area due to 
lateral contraction during tensile loading, the constant-volume 
hypothesis was applied~\cite{Dieter1988,Choung2008}. In the present 
simulations the $z$-direction is a free surface with vacuum above and 
below the slab; as deformation proceeds the free surface becomes 
increasingly wavy, particularly at larger grain sizes where necking 
develops, so the $z$ box dimension cannot be used to estimate the 
true load-bearing cross-sectional area. Only the $x$ and $y$ box 
dimensions are therefore used. Under uniaxial tensile loading along 
the $x$-axis, the box expands in the $x$-direction and contracts in 
the $y$-direction. The geometric correction factor at strain 
$\varepsilon$ is computed as
\begin{equation}
    \label{eq:f}
    f(\varepsilon) = \frac{x(\varepsilon)\, y(\varepsilon)}{x_0\, y_0},
\end{equation}
where $x_0$, $y_0$ are the box dimensions at zero strain and 
$x(\varepsilon)$, $y(\varepsilon)$ are the box dimensions at the 
strain at which the flow stress is extracted. The corrected flow 
stress is then
\begin{equation}
    \label{eq:sigma_corrected}
    \sigma_{\text{corrected}} = \sigma_{\text{eng}}(\varepsilon) f(\varepsilon),
\end{equation}

This approach accounts directly for the actual lateral contraction of 
the simulation box at the point of flow stress extraction, without 
assuming a uniform strain distribution across the slab thickness, 
making it more physically appropriate for samples undergoing 
localized necking.

\begin{figure}[ht!]
	\centering
	\includegraphics[height=5.5cm]{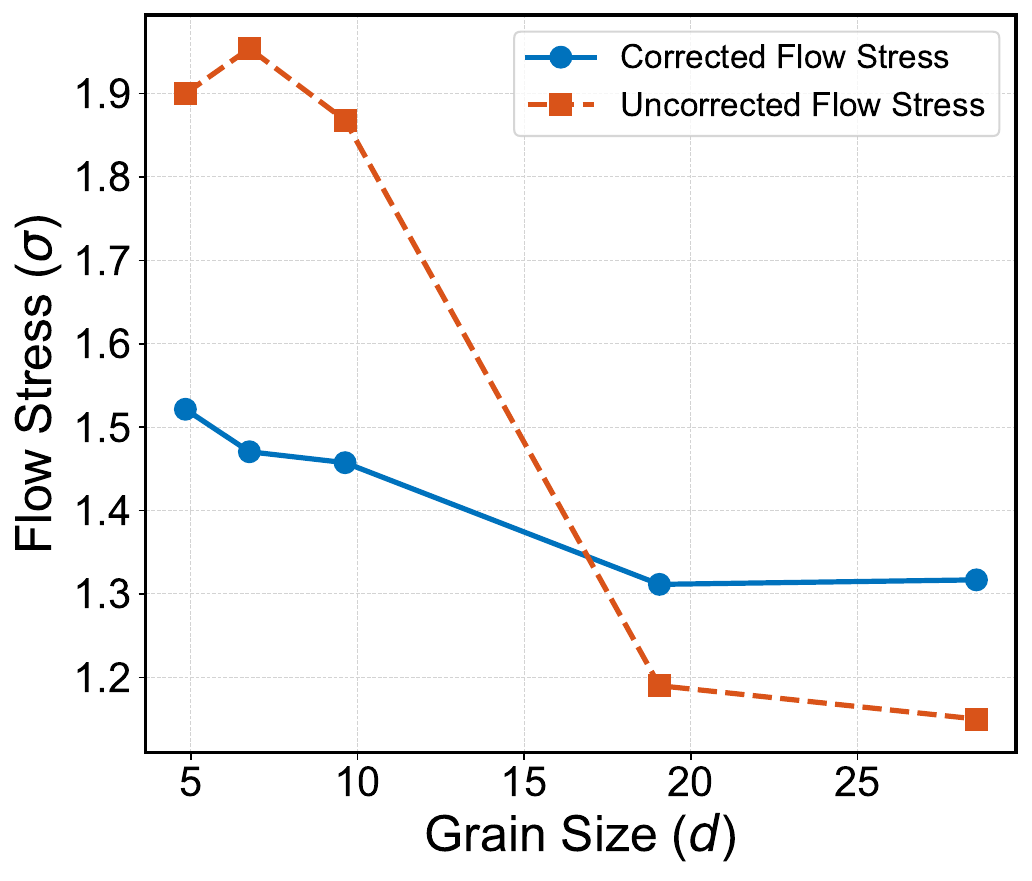}
    \caption{Comparison of corrected and uncorrected flow stress as a function of mean grain size $d$ for cylindrical sintered (CS) samples simulated using the \textit{Pascuet15}~\cite{PASCUET2015229} MEAM potential. The correction accounts for the lateral contraction of the simulation box using the constant-volume hypothesis, applied at the strain at which the flow stress is extracted. The difference between the two curves is most pronounced for the larger grain sizes, where necking introduces geometric softening.}
	\label{fig:flowstress_corrected}
\end{figure}

The corrected and uncorrected flow stress values as a function of 
grain size are presented in Fig.~\ref{fig:flowstress_corrected}. For 
the smaller grain sizes ($d \leq 9.63\,\mathrm{nm}$), where plastic 
flow is stable and necking is absent, the correction is moderate, with 
corrected values of 1.46--1.52\,GPa compared to uncorrected values of 
1.87--1.95\,GPa. For the larger grain sizes ($d = 19.06$ and 
$28.57\,\mathrm{nm}$), the correction has a more significant effect: 
the uncorrected flow stress drops sharply to approximately 1.19 and 
1.15\,GPa, respectively, due to geometric softening caused by necking, 
while the corrected values of 1.31 and 1.32\,GPa are higher and more 
representative of the true plastic resistance of the material.

Wang~\textit{et al.}~\cite{Wang2025} reported a transition from inverse Hall--Petch to Hall--Petch behavior in nanocrystalline Al at a critical grain size of 10--12\,nm for a two-dimensional equiaxial grain structure. In the present study, no such transition is observed within the grain-size range investigated (4.84--40.34\,nm). The corrected flow stress continues to decrease monotonically with increasing grain size, and the magnitude of this decrease is consistent with inverse Hall--Petch behavior throughout. The apparent flattening of the corrected flow stress curve at larger grain sizes does not constitute evidence of a Hall--Petch transition, as the decrease remains gradual and the values do not show the increasing trend with grain size that would be characteristic of conventional dislocation-mediated strengthening. It should furthermore be noted that for the larger grain sizes ($d = 19.06$ and $28.57\,\mathrm{nm}$), the extraction of a reliable flow stress is compromised by the onset of necking and geometric instability, making it difficult to draw definitive conclusions about the flow stress behavior in this regime. A reliable determination of whether a Hall--Petch transition occurs at larger grain sizes would require either larger simulation boxes that suppress geometric instability or alternative loading protocols that avoid the onset of necking.

\subsection{Deformation mechanisms}
\label{sec:defmech}

\begin{figure}[ht!]
	\centering
	\begin{subfigure}{0.48\columnwidth}
		\centering
		\includegraphics[width=\textwidth]{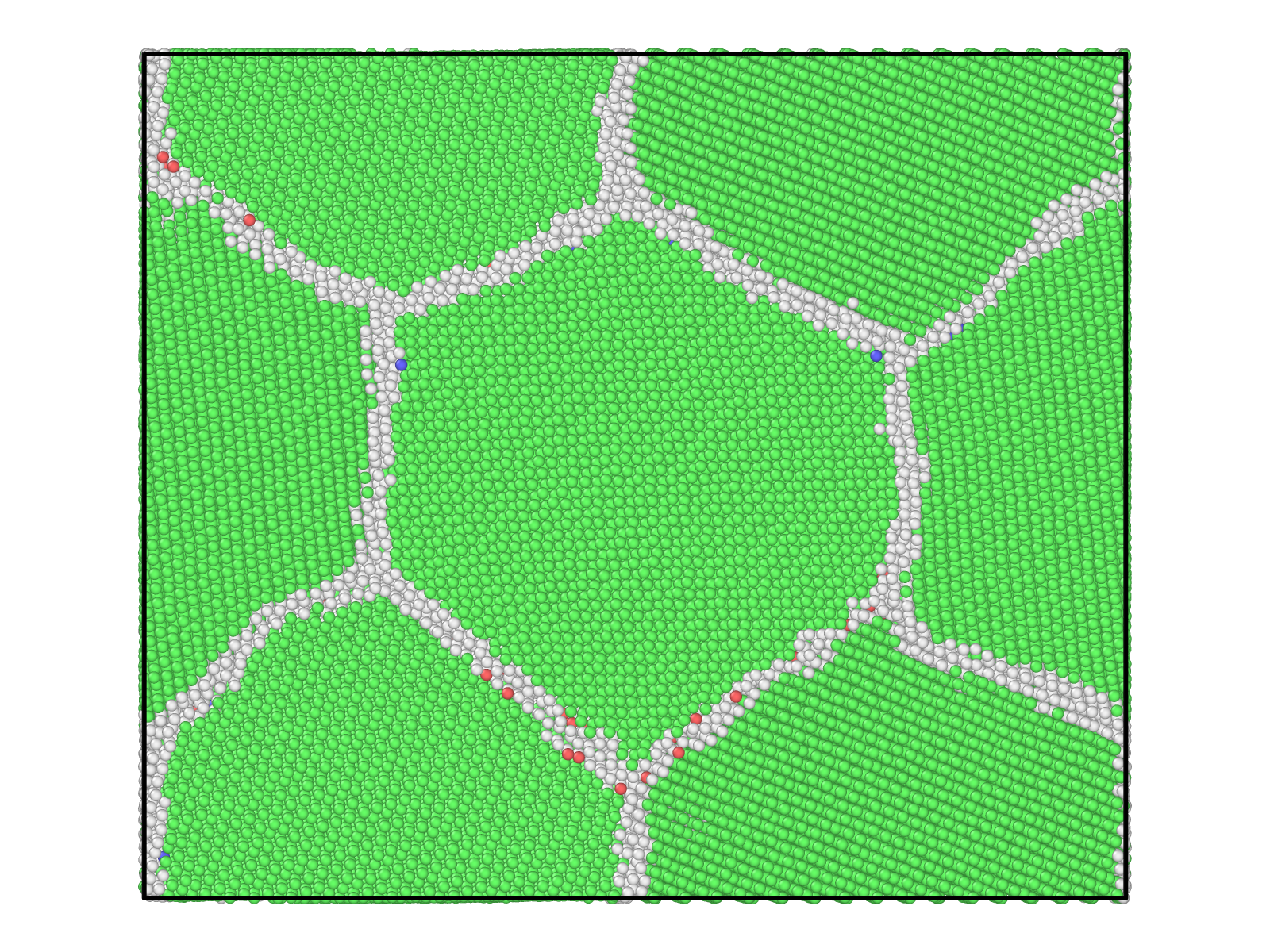}
		\caption{0\% CNA}
	\end{subfigure}\hfill
	\begin{subfigure}{0.48\columnwidth}
		\centering
		\includegraphics[width=\textwidth]{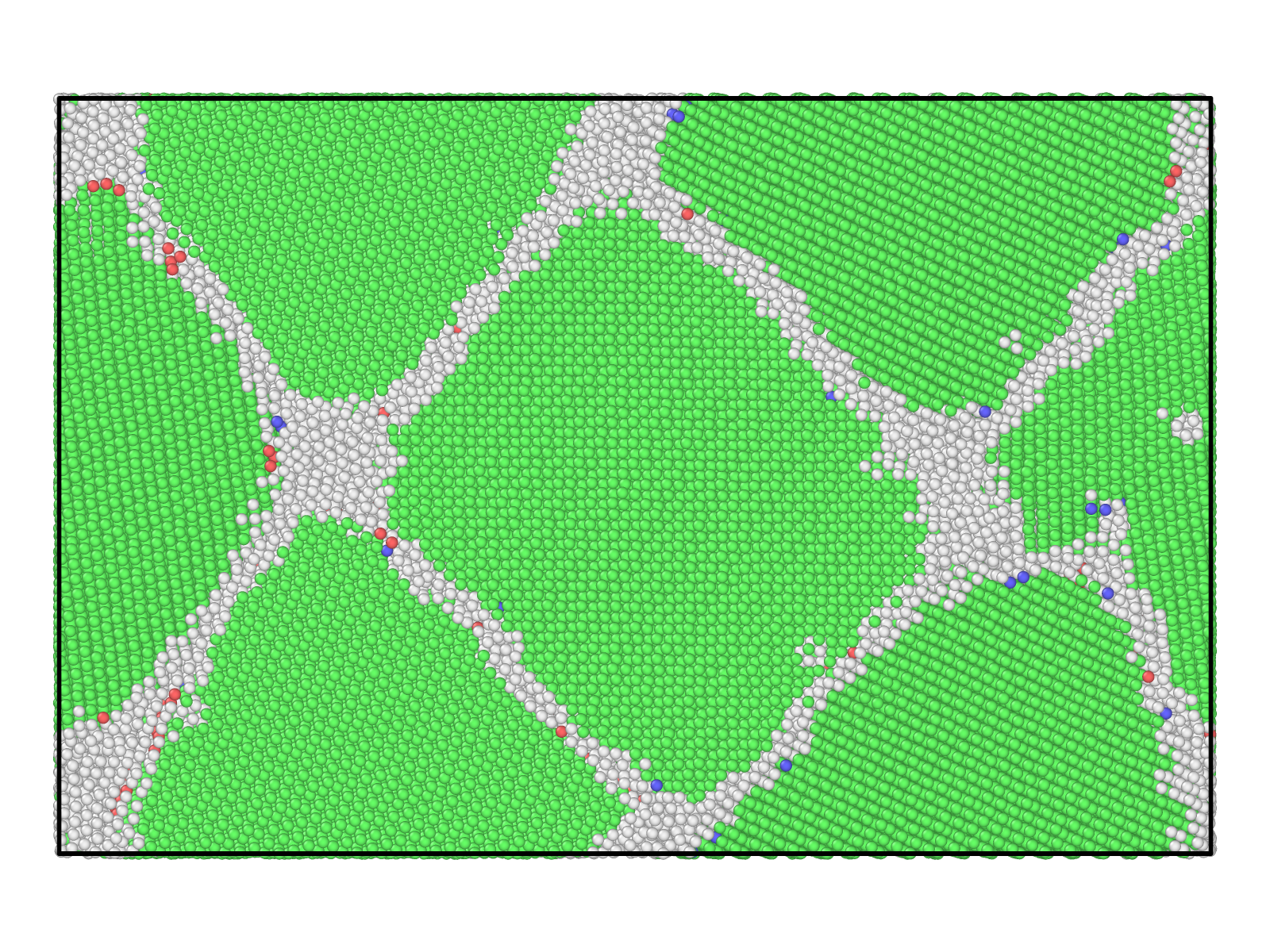}
		\caption{16\% CNA}
	\end{subfigure}
	
	\caption{Microstructural evolution of the cylindrical sintered (CS) sample with an average grain size of 9.63\,nm and fully random crystallographic orientations along all three axes, at 0\% and 16\% tensile strain, simulated using the \textit{Pascuet15}~\cite{PASCUET2015229} MEAM potential and analyzed using OVITO~\cite{Stukowski2012} via Common Neighbor Analysis (CNA).}
	\label{fig:cna_dxa_all}
\end{figure}

Fig.~\ref{fig:cna_dxa_all} presents the microstructural evolution of the cylindrical sintered (CS) sample with a grain size of 9.63\,nm at 0\% and 16\% tensile strain, constructed with fully random crystallographic orientations along all three axes and simulated using the \textit{Pascuet15}~\cite{PASCUET2015229} MEAM potential. The microstructural evolution is visualized using Common Neighbor Analysis (CNA), as implemented in OVITO~\cite{Stukowski2012}.

A general observation concerns the interpretation of the Dislocation Extraction Algorithm (DXA) at grain boundaries. DXA identifies dislocation segments from the deviation of atomic neighborhoods from the ideal fcc structure. At grain boundaries, however, the atomic arrangement is intrinsically disordered, and this disorder produces signals that DXA attributes to dislocation segments even when no plasticity-carrying dislocations are present. The dislocation densities and dislocation types reported by DXA within the grain-boundary regions should therefore not be interpreted as a quantitative measure of plastic activity. In both CS and HV samples investigated in this work, essentially no dislocation segments are detected inside the grain interiors throughout the deformation, confirming that plasticity in these nanocrystalline samples is mediated by grain-boundary processes rather than by intragranular dislocation glide.


At 0\% strain, the grain boundaries are relatively narrow and well-defined, with only isolated defects at triple junctions and no dislocations present within the grain interiors. By 16\% strain, the grain boundaries have broadened considerably and the grain morphology has evolved noticeably under the imposed uniaxial tension. Since the CS samples are cut as cylindrical grains from an underlying hexagonal Voronoi tessellation, each grain begins with an approximately hexagonal cross-section. During straining, two of the six grain boundaries of each grain progressively shrink, while the remaining four extend, so that the initially hexagonal cross-section evolves into a roughly square shape at large strains. This grain-shape evolution is a direct geometric consequence of the imposed uniaxial loading and is observed consistently across the CS samples for both interatomic potentials investigated.

Unlike the 9.63\,nm sample, dislocations are present within the grain interiors already at 0\% strain. This is a consequence of the sintering protocol: the sintered sample was subjected to a compaction step at 93\% of its equilibrium volume under an NVT ensemble to eliminate residual voids prior to tensile loading. In the smaller 9.63\,nm sample, the higher grain-boundary area-to-volume ratio provides sufficient sink density for these compaction-induced dislocations to be absorbed by the surrounding grain boundaries during the NVT relaxation. In the larger 19.06\,nm sample, the greater distance between grain boundaries and the lower sink density means that not all intragranular dislocations escape within the relaxation time, leaving a residual intragranular dislocation content at the start of the tensile simulation.

Despite this difference in the initial state, the overall deformation morphology follows the same trend as for the smaller sample: the hexagonal grain cross-sections evolve toward roughly square shapes, and the residual intragranular dislocations progressively diminish during straining, having essentially vanished by 16\% strain through migration to the grain boundaries or free surfaces. Their disappearance during the early stages of deformation suggests that their presence does not significantly influence the macroscopic mechanical response, and the dominant deformation mechanism remains grain-boundary-mediated plasticity, consistent with the smaller sample.

\section{Conclusions}
\label{sec:conclusions}

This work has introduced the cylindrical sintering (CS) method for constructing nanocrystalline aluminum thin-film samples with fully random three-dimensional grain orientations, and benchmarked it against a sintered hexagonal Voronoi (HV) reference under identical sintering and equilibration conditions using two qualitatively different interatomic potentials, the classical \textit{Pascuet15} MEAM potential and the \textit{tabGAP} machine-learning potential. The principal findings are as follows.

\begin{itemize}
	\item \textbf{Cylindrical sintering produces wider and softer grain boundaries than sintered Voronoi tessellation.} The two-phase model parameters show that the CS samples have a softer grain interior and a markedly softer grain boundary than the HV samples for both potentials, even though the fitted grain-boundary thickness is smaller for CS than for HV. This effect is reproduced consistently with both potentials, confirming that it is a genuine consequence of the sintering method rather than an artifact of any particular interatomic description.
	
	\item \textbf{Inverse Hall--Petch behavior across the full grain-size range.} The elastic modulus, ultimate tensile strength, and engineering yield stress all decrease monotonically with decreasing grain size from 40.34 down to 4.84\,nm for both CS and HV samples and for both potentials, with no transition to conventional Hall--Petch strengthening observed within this range.
	
	\item \textbf{The sample-construction method and the interatomic potential contribute additive, largely independent offsets.} Switching from HV to CS shifts the predicted mechanical properties by a comparable margin across both potentials, and switching from \textit{Pascuet15} to \textit{tabGAP} shifts them by a similar fraction across both sample geometries. Disentangling the two effects requires the kind of fully crossed potential-by-method study performed here.
\end{itemize}

The cylindrical sintering method provides a route to physically realistic, geometrically controlled nanocrystalline samples that bridges the gap between idealized Voronoi geometries and the more disordered boundaries produced by melt--cool methods, and is particularly suited to MD studies of columnar thin-film microstructures in which deterministic grain geometry is required alongside realistic grain-boundary structure.

\section*{Declaration of Competing Interest}
The authors declare that they have no known competing financial interests or personal relationships that could have appeared to influence the work reported in this paper.

\section*{Acknowledgments} 
This work was supported by the Ministry of Education, Youth and Sports of the Czech Republic under the Operational Programme Johannes Amos Comenius (OP JAK), project No. CZ.02.01.01/00/23\_020/0008549 (PSSITE). The authors also acknowledge financial support from the INTER-COST project (grant no. LUC24093). Computational resources were provided by the e-INFRA CZ project (ID: 90254). The INTER-COST and e-INFRA CZ projects are supported by the Ministry of Education, Youth and Sports of the Czech Republic.

\section*{Data availability}
Data will be made available on request.


\bibliography{main.bib}

\end{document}